\documentclass[lettersize,journal]{IEEEtran}
\usepackage{amsmath,amsfonts}
\usepackage[utf8]{inputenc}
\usepackage{algorithm}
\usepackage{algpseudocode}
\usepackage{array}
\usepackage[caption=false,font=normalsize,labelfont=sf,textfont=sf]{subfig}
\usepackage{textcomp}
\usepackage{stfloats}
\usepackage{url}
\usepackage{verbatim}
\usepackage{graphicx}
\usepackage{cite}
\usepackage{multirow}
\usepackage{soul}
\usepackage{xcolor}
\usepackage{booktabs}
\begin{document}

\title{Turning Noise into Value: Uncovering Service Preferences from Ambiguous Interaction in E-commerce}

\author{Cheng Li,~\IEEEmembership{Graduate Student Member,~IEEE,}~Yong Xu,~Suhua Tang,~\IEEEmembership{Senior Member,~IEEE,}\\~Wenqiang Lin,~Xin He,~\IEEEmembership{Member,~IEEE,}~Jinde Cao,~\IEEEmembership{Fellow,~IEEE}
\thanks{This work was supported in part by National Natural Science Foundation of China under Grant 62072004 and 62576098. (\textit{Corresponding authors: Yong Xu and Suhua Tang.})}
\thanks{Cheng Li, Yong Xu, Wenqiang Lin, and Xin He, are with the School of Computer and Information, Anhui Normal University, Wuhu, Anhui  241002, China (e-mail: jxxylc@ahnu.edu.cn; yxull@ahnu.edu.cn; wenqiang\_lin@ahnu.edu.cn; xin.he@ahnu.edu.cn).

Suhua Tang is with the Department of Computer and Network Engineering, Graduate School of Informatics and Engineering, The University of Electro Communications, Tokyo 182-8585, Japan (e-mail: shtang@uec.ac.jp).

Jinde Cao is with the School of Mathematics Southeast University, Nanjing 211189, China, and also with the Purple Mountain Laboratories, Nanjing 211111, China (email: jdcao@seu.edu.cn).}}%

\maketitle

\begin{abstract}
In e-commerce service recommendation, utilizing auxiliary behaviors to alleviate data sparsity often relies on the flawed assumption that auxiliary behaviors that fail to trigger target actions are negative samples. This approach is fundamentally flawed as it ignores false negatives where users actually harbor latent intent or interest but have not yet converted due to external factors. Consequently, existing methods suffer from sample selection bias and a severe distribution shift between the auxiliary and target behaviors, leading to the erroneous suppression of potential user needs. To address these challenges, we propose a Noise-to-Value Adapter (NoVa), an e-commerce service recommendation framework that re-examines the problem through the lens of positive-unlabeled learning. Instead of treating ambiguous auxiliary behaviors as definite negatives, NoVa aims to uncover high-quality preferences from noise via two key mechanisms. First, to bridge the distribution gap, we employ adversarial feature alignment. This module aligns the auxiliary behavior distribution with the target space to identify high-confidence false negatives, which are instances that statistically resemble confirmed target behaviors and thus represent latent conversion intents. Second, to mitigate label noise caused by accidental clicks or random browsing, we introduce a semantic consistency constraint. This mechanism implements semantic-aware filtering based on the content similarity of services, acting as a bias correction step to filter out low-confidence interactions that lack semantic relevance to historical user preferences. Extensive experiments on three real-world datasets demonstrate that NoVa outperforms state-of-the-art baselines.
\end{abstract}


\section{Introduction}
\IEEEPARstart{I}{n} the rapidly expanding ecosystem of web services, recommender systems have become a core technology in platforms such as e-commerce and social media by uncovering users’ potential interests through mining their historical interactions~\cite{r1,r2}. Particularly in e-commerce service platforms, the primary objective is to accurately predict user invocations for target services (e.g., purchasing a product or subscribing to a plan), which directly drive platform revenue~\cite{d1}. Collaborative filtering (CF) is one of the most widely adopted paradigms, which learns latent representations of users and items by analyzing the implicit associations in the user-item interaction matrix to enable personalized recommendation~\cite{r3,r4}. This paradigm has demonstrated remarkable effectiveness in various web applications by capturing the collaborative patterns among users and services~\cite{d3}. However, traditional CF methods typically rely on modeling a single type of user behavior, which limits their performance in scenarios such as user cold-start and data sparsity~\cite{b1}. 

With the continuous expansion of recommendation scenarios, user-item interactions have become increasingly diverse~\cite{d2}. Modeling user preferences based solely on a single behavior type is no longer sufficient to fully capture the complexity of user interests~\cite{b2,b3}. In practice, users often engage in multiple types of interactions (e.g., view, favorite, cart\footnote{In this article, ``cart'' and ``add-to-cart'' are equivalent.}, purchase), which provide rich, multi-dimensional information for preference modeling~\cite{c1}. These diverse behaviors also reflect underlying patterns of interest shifts and behavioral transitions~\cite{b4}. Multi-behavior recommendation leverages the dependencies and complementary information among different behavior types to enhance the system’s ability to model the diversity of user interests~\cite{r9,r15}. This, in turn, effectively alleviates longstanding challenges faced by collaborative filtering methods, such as interaction sparsity and the cold-start problem~\cite{r10}. By transferring knowledge from abundant auxiliary behaviors to sparse target behaviors, these methods aim to achieve a more comprehensive and accurate prediction of user service demands~\cite{d4}.

\begin{figure}[t]
\centering
\includegraphics[width=0.95\columnwidth]{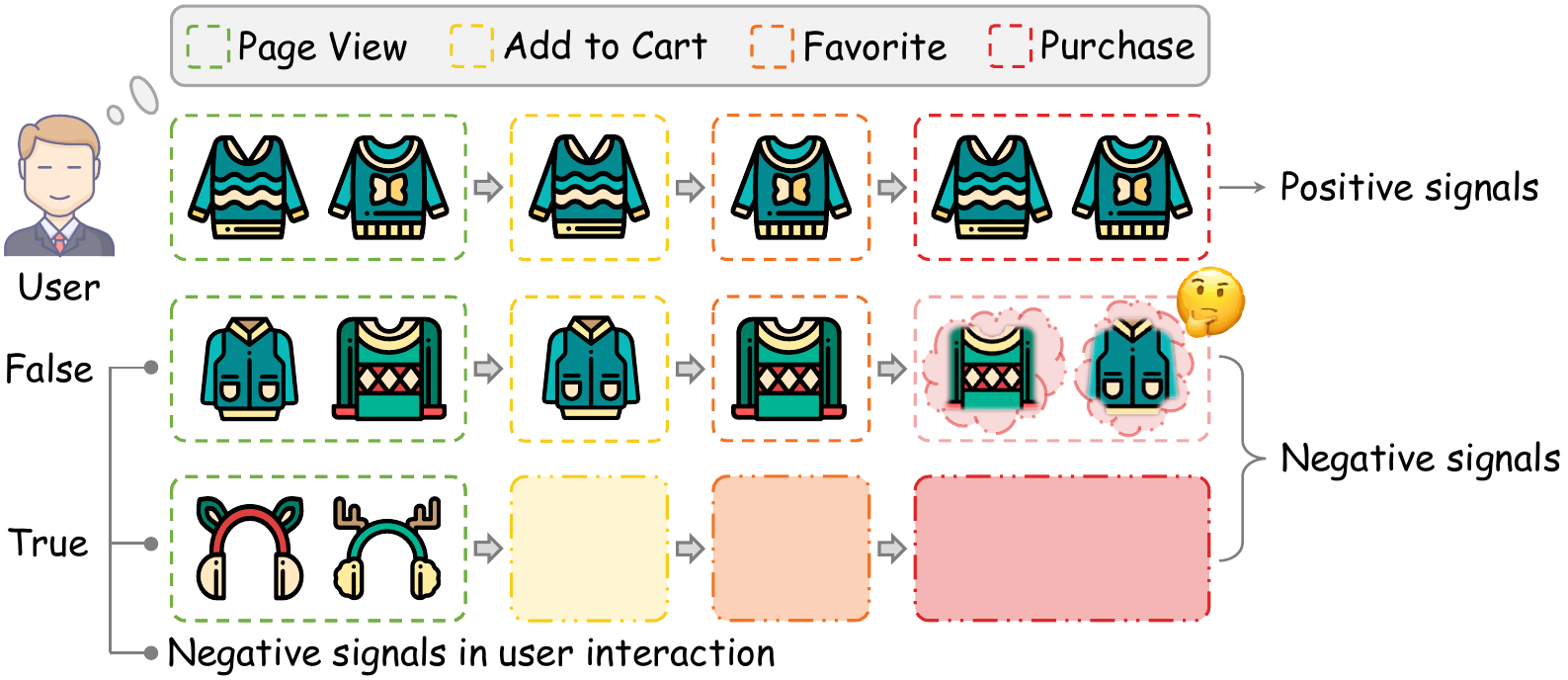} 
\caption{An illustrative example of the positive-unlabeled nature in service interactions, distinguishing high-confidence false negatives from label noise.}
\label{fig1}
\end{figure}

However, despite the significant progress of multi-behavior recommendation in alleviating data sparsity and modeling user preferences~\cite{r11,r14}, existing methods typically operate under a closed-world assumption~\cite{d5}: they indiscriminately treat all auxiliary interactions that do not trigger a target invocation as negative samples. This approach is fundamentally flawed as it ignores the positive-unlabeled nature of implicit service feedback~\cite{d6}, failing to discriminate the true semantics of non-converted interactions. As illustrated in Fig.~\ref{fig1}, this naive assumption introduces two critical issues rooted in the data distribution. First, it overlooks false negatives, where users possess latent intent or interest in a service but have not yet converted due to external factors (e.g., comparison or delay) rather than dislike~\cite{r12}. Ignoring these potential positives leads to sample selection bias, causing the model to erroneously suppress users' true needs. Second, it neglects the significant distribution shift between exploratory auxiliary behaviors (e.g., random browsing) and decisive target behaviors~\cite{d8}. Some auxiliary interactions are merely label noise resulting from accidental clicks or aimless exploration, which deviate significantly from the target preference distribution~\cite{d7}. Incorporating such heterogeneous signals indiscriminately into preference modeling leads to negative transfer effects, where the model learns biased patterns from irrelevant noise~\cite{r13}. Therefore, effectively uncovering high-confidence false negatives to recover latent intents, while simultaneously suppressing label noise caused by distribution shifts, remains a crucial research challenge for enhancing the quality of recommendation.

To address the above issues, we propose a \textbf{No}ise-to-\textbf{Va}lue Adapter (NoVa). Grounded in positive-unlabeled learning, NoVa focuses on recovering high-confidence false negatives and suppressing label noise. Specifically, to uncover latent intents hidden within sparse data, we treat the identification of potential needs as a distribution alignment problem. We first construct view-specific subgraphs for auxiliary and target behaviors. Through adversarial feature alignment on these subgraphs, the feature network of auxiliary behavior is trained to align with the target behavior distribution. Based on this alignment, auxiliary behaviors that do not produce target invocations but exhibit high structural similarity to target samples are identified as high-confidence false negatives (i.e., latent positive samples) and integrated into the main recommendation task. For interaction noise, we introduce a semantic consistency constraint. Instead of indiscriminate fusion, we guide the feature transfer process via service content similarity. By calculating the semantic deviation between auxiliary interactions and historical target preferences, label noise (e.g., accidental clicks) is effectively suppressed and filtered within the multi-head attention-based feature transfer. 

In general, the main contributions of this paper are as follows:
\begin{itemize}
    \item We innovate the multi-behavior service recommendation paradigm by formulating it within the positive-unlabeled learning framework. Specifically, we argue that non-converted service logs are not merely negative samples but often contain valuable false negatives due to distribution shifts. Based on this insight, we propose an adversarial alignment mechanism to bridge the distribution gap and recover latent user needs.
    \item We propose a semantic-aware filtering mechanism to tackle the label noise inherent in web service logs. By enforcing semantic consistency constraints based on item similarity, NoVa effectively distinguishes random noise from genuine interests. This approach mitigates the negative transfer effects caused by the inclusion of irrelevant auxiliary behaviors.
    \item We validate the robustness and effectiveness of the proposed method on three real-world e-commerce datasets. Experimental results demonstrate that NoVa consistently outperforms state-of-the-art methods on HR@10 and NDCG@10.
\end{itemize}
\section{Related Work}

\subsection{Multi-Behavior Recommendation}
By incorporating auxiliary behaviors, multi-behavior recommendation enhances the modeling of users' target behaviors and has achieved success in alleviating data sparsity and improving recommendation performance~\cite{c2}. Early approaches primarily relied on matrix factorization or sampling strategies to integrate multi-behavior features. For example, BF~\cite{r16} decomposed different types of behavioral signals into independent latent representations to model user interests. With the rapid development of graph convolutional networks (GCN), an increasing number of methods have adopted graph-based architectures to capture cross-behavior dependencies and collaborative patterns. For instance, MATN~\cite{r5} aligns multiple types of user interactions to capture the intrinsic correlations between behaviors and employs an attention mechanism to focus on behavior features that contribute more to the prediction of the target behavior. To address the sparsity of target behavior signals, CML~\cite{r17} formulates a multi-behavior contrastive learning task to promote knowledge transfer across behaviors and introduces a contrastive meta-network to model the heterogeneity of user preferences.

In recent years, increasing attention has been paid to the hierarchical and cascading relationships among various user behaviors. For instance, NMTR~\cite{r7} jointly models user behaviors through multi-task learning and explicitly captures the cascade relationship (e.g., view$\to$cart$\to$purchase) to enforce the sequential dependencies of user actions. MB-CGCN~\cite{r18} further extends this by modeling real-world behavior chains via graph convolution to capture higher-order sequential dependencies. To incorporate temporal dynamics, KHGT~\cite{r8} integrates a hierarchical graph structure with temporal encoding to capture the evolving fine-grained user preferences. Additionally, BCIPM~\cite{r19} learns item-specific user preferences under each behavioral context separately and selectively retains those preference representations relevant to the target behavior during the final recommendation stage. However, despite these advancements, most existing methods typically operate under a closed-world assumption: they indiscriminately treat all auxiliary interactions that do not trigger a target invocation as negative samples. This approach is fundamentally flawed as it overlooks the positive-unlabeled nature of service interaction logs. Specifically, these methods fail to distinguish between false negatives (latent user needs that have not yet converted) and true negatives (noise or dislike). Consequently, they neglect the significant distribution shift between exploratory auxiliary behaviors and decisive target behaviors. Directly transferring features from auxiliary domains without aligning these distributions or filtering noise leads to sample selection bias, limiting the model's robustness in capturing true user service demands.

\subsection{Multi-Task Recommendation}
Recently, some studies have attempted to integrate behavior feature modeling with multi-task learning to better leverage auxiliary behaviors for enhanced user modeling. For example, CRGCN~\cite{r20} employs a cascaded residual GCN to refine user and item embeddings, establishing connections between early and later behaviors during the embedding learning process. To address data sparsity, RCL~\cite{r14} proposes a relation-aware contrastive learning framework that explicitly models the complex dependencies among heterogeneous behavioral relations, ensuring robust representation learning under sparse target signals. Incorporating external knowledge, KMCLR~\cite{r26} utilizes knowledge graph semantics to align multi-behavior signals, thereby enhancing the semantic richness of item representations. Furthermore, AITM~\cite{d9} proposes an adaptive information transfer module to model the sequential dependence between the auxiliary task and the target task, aiming to estimate the conversion probability more accurately. 

Despite these mechanisms, challenges remain in handling noisy and complex interactions. MPC~\cite{r13} constructs a negative graph to identify and filter out misleading user–item interactions, while COPF~\cite{r21} formulates multi-behavior fusion as a combinatorial optimization problem to mitigate negative transfer caused by inconsistent distributions. However, while these methods attempt to reduce noise or optimize fusion, they primarily focus on feature-level correlation or overlap. They largely overlook the distribution shift between the auxiliary and target behaviors, failing to explicitly align the underlying feature distributions. More critically, from a positive-unlabeled learning perspective, these approaches typically treat non-converted interactions as hard negatives or simple noise. Consequently, they lack a mechanism to recover high-confidence false negatives, which represent latent service needs that are structurally similar to target samples but masked by distribution discrepancies. This limitation restricts their ability to fully exploit the potential value of ambiguous auxiliary logs in sparse service recommendation scenarios.

\subsection{Positive-Unlabeled Learning in Recommendation}
In web service recommendation, user feedback is predominantly implicit (e.g., service invocations), creating a scenario where only positive samples are observed while explicit negative samples are unavailable. Such a setting inherently constitutes a positive-unlabeled learning problem, where the unobserved interactions are not merely negatives but a mixture of actual negative preferences (true negatives) and latent interests that have not yet been discovered (false negatives)~\cite{d10}. Several representative models have been proposed to address this challenge. For instance, OCCF~\cite{d11} was among the first to formulate this problem, treating unobserved entries as negative samples with lower confidence weights rather than absolute negatives. To capture more complex data distributions, IRGAN~\cite{d12} adopted a generative adversarial network (GAN) approach, where a generator creates difficult negative samples to confuse the discriminator, implicitly distinguishing between true negatives and potential positives. 

More recently, researchers have focused on refining the sampling strategy within the positive-unlabeled framework to better distinguish latent positive samples from the vast unlabeled data. SRNS~\cite{d13} utilizes the variance of prediction scores as a measure of uncertainty, explicitly identifying items with high variance as potential false negatives rather than treating them as definite negative samples. Similarly, to overcome the limitations of random sampling in the unlabeled space, MixGCF~\cite{d14} proposes a hop-mixing technique to synthesize hard negative samples. This approach implicitly navigates the boundary between positive and unlabeled data, ensuring that the model learns discriminative features without being misled by the false negative nature of unobserved interactions. However, these existing positive-unlabeled learning models are primarily designed for single-behavior contexts. They operate under the assumption that the unobserved data follows a consistent distribution within a single domain. In our multi-behavior service ecosystem, auxiliary behaviors (e.g., exploratory browsing) and target behaviors (e.g., decisive purchasing) follow distinct statistical distributions. Consequently, they struggle to recover high-confidence false negatives that are structurally similar to target samples but masked by the distribution shift between behaviors.

\section{Problem Definition}
To enhance clarity, we first introduce the notations and task definition used in this paper. Let the set of user behavior types be denoted as $\mathcal{B} = \{b_1, b_2, \cdots, b_K\}$, where $b_k$ represents the $k$-th type of behavior (e.g., click, cart), and $b_{k^*}$ denotes the target behavior (e.g., purchase). Based on this, we construct $K$ user–item bipartite graphs $\mathcal{G} = \{\mathcal{G}_1, \mathcal{G}_2, \cdots, \mathcal{G}_K\}$, where each subgraph $\mathcal{G}_k = (\mathcal{V}_k, \mathcal{E}_k)$ corresponds to interactions under behavior $b_k$. The node set is defined as $\mathcal{V}_k = \mathcal{U}_k \cup \mathcal{I}_k$, where $\mathcal{U}_k$ is the set of users and $\mathcal{I}_k$ is the set of items. The edge set $\mathcal{E}_k$ contains user–item interaction edges under behavior $b_k$. Accordingly, the interaction matrix for behavior $b_k$ is represented as $\mathbf{B}_{k}=\begin{bmatrix}{b_{k,ui}}\end{bmatrix}_{|\mathcal{U}|\times|\mathcal{I}|}\in{\{}0,1\}$, where $b_{k,ui} = 1$ indicates that user $u$ has interacted with item $i$ under behavior $b_k$, and $b_{k,ui} = 0$ otherwise. Let $\mathcal{U}=\cup_{k=1}^{K}\mathcal{U}_{k}$ and $\mathcal{I}=\cup_{k=1}^{K}\mathcal{I}_{k}$ denote the complete sets of \(M\) users and \(N\) items, respectively. The interactions across all behaviors form a multi-behavior graph $\mathcal{G}=(\mathcal{V},\mathcal{E},\mathcal{B})$, where $\mathcal{V}=\mathcal{U} \cup \mathcal{I}$ and $\mathcal{E}=\cup_{k=1}^{K}\mathcal{E}_{k}$.

Crucially, for the target behavior $b_{k^*}$, the interaction data exhibits a positive-unlabeled nature. The observed interaction set $\mathcal{O}^+ = \{(u,i) \mid b_{k^*, ui} = 1\}$ represents positive samples (confirmed target behaviors). However, the unobserved set $\mathcal{O}^? = \{(u,i) \mid b_{k^*, ui} = 0\}$ is unlabeled, which theoretically consists of a mixture of true negatives and false negatives. Traditional methods typically assume $\mathcal{O}^? \approx negative$, leading to sample selection bias. Therefore, our objective is to learn a predictive function that can distinguish latent positive signals from the unlabeled set $\mathcal{O}^?$ by leveraging the structural correlations in auxiliary graphs $\mathcal{G}_{k \neq k^*}$, ultimately estimating the true probability $y_{u,i}^{k^*}$ that user $u$ will perform the target behavior on item $i$.

\section{Methodology}
\begin{figure*}[t]
\centering
\includegraphics[width=0.95\linewidth]{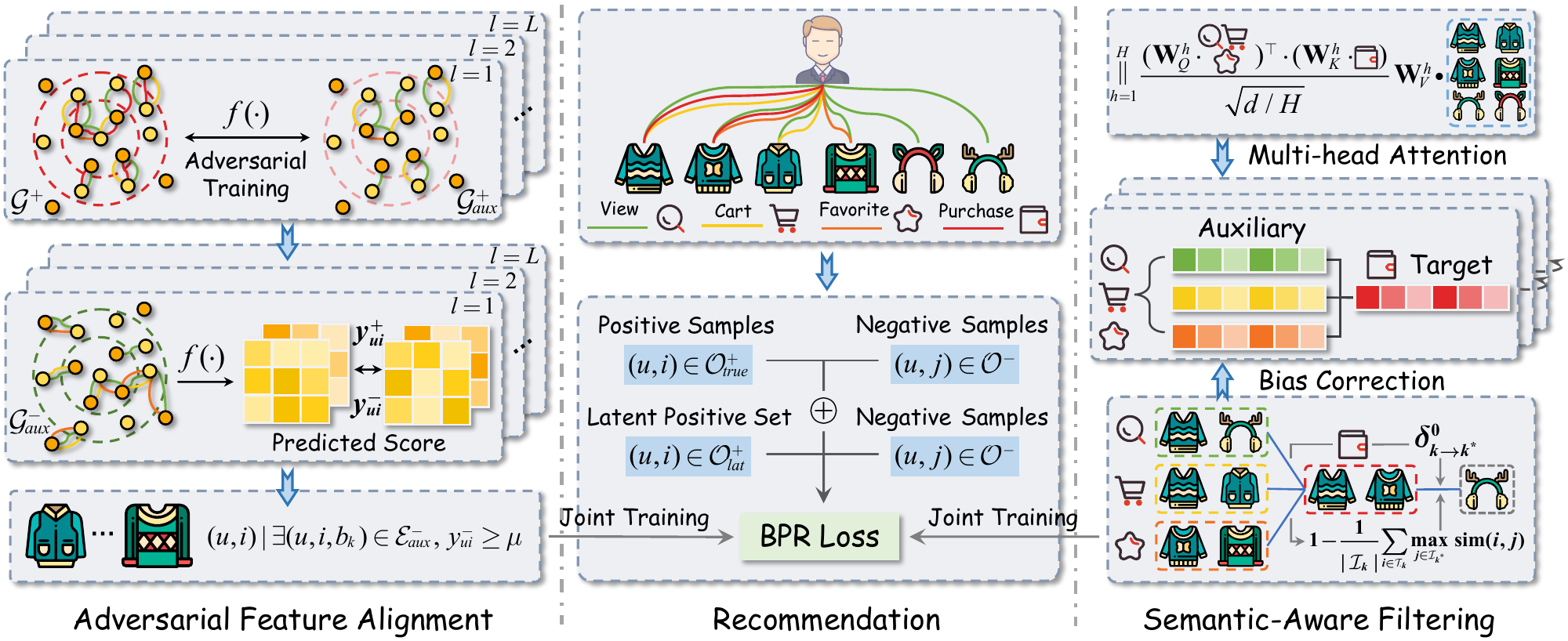} 
\caption{The overall framework of the proposed NoVa model. Grounded in a positive-unlabeled learning perspective, NoVa consists of two core modules: (1) adversarial feature alignment for bridging distribution gaps and recovering high-confidence false negatives; and (2) semantic-aware filtering for suppressing label noise via consistency constraints. The refined value signals extracted by these modules are then integrated into the target service recommendation task.}
\label{fig2}
\end{figure*}
\subsection{Overview}
Fig.~\ref{fig2} illustrates the overall framework of NoVa, which is designed to tackle the inherent challenges of distribution shift and label noise in multi-behavior service recommendation. Specifically, raw auxiliary interaction logs are a mixture of valuable signals and irrelevant noise. To disentangle them, NoVa incorporates two complementary components that operate on different modeling levels. The first component, the adversarial feature alignment module, functions at the interaction level. Recognizing that the distribution of auxiliary behaviors differs from that of target behaviors, this module constructs view-specific subgraphs and employs adversarial learning to bridge the distribution gap. This process enables the identification of high-confidence false negatives, which represent interactions that statistically resemble target behaviors and thus imply latent conversion intents. In contrast, the second component, the semantic-aware filtering module, operates at the representation level. It addresses the issue of label noise that can lead to negative transfer. By leveraging service content similarity and enforcing semantic consistency constraints, this module effectively filters out misleading features during multi-behavior fusion. 

\subsection{False Negative Discovery}
\subsubsection{View-Specific Subgraph Construction}
In real-world service scenarios, users often engage in frequent auxiliary behaviors without immediately invoking the target service. From a positive-unlabeled learning perspective, treating these non-converted interactions indiscriminately as negative samples is flawed. Instead, they often represent false negatives—latent service needs that have not yet translated into final actions due to external factors. However, a significant distribution shift typically exists between exploratory auxiliary behaviors and decisive target invocations. Directly utilizing auxiliary features without alignment may introduce bias, limiting the model's ability to accurately capture the true distribution of user preferences. To bridge this gap and recover high-confidence false negatives, we first need to establish a supervised signal that captures the correlation between auxiliary and target distributions. We construct a positive context subgraph $\mathcal{G}^+$, which contains all historical interactions for user-service pairs where the target behavior has definitively occurred. The edge set is formally defined as:
\begin{equation}
\mathcal{E}^+ = \{(u, i, b_k)\mid (u, i, b_{k^*}) \in \mathcal{E}_{k^*},b_k \in \mathcal{B}\},
\end{equation}
where $\mathcal{E}_{k^*}$ denotes the edge set of the target behavior subgraph $\mathcal{G}_{k^*}$. Essentially, $\mathcal{G}^+$ represents the complete behavioral footprint of successful service invocations. To explicitly model the feature mapping from the auxiliary domain to the target domain, we separate the auxiliary components from this positive context. We derive an auxiliary-specific subgraph $\mathcal{G}_{aux}^+=(\mathcal{V}_{aux}^+,\mathcal{E}_{aux}^+)$ by retaining only the auxiliary edges within $\mathcal{G}^+$. Its edge set is defined as:
\begin{equation}
\mathcal{E}_{aux}^+ = \{(u, i, b_k)\mid (u, i, b_{k^*}) \in \mathcal{E}_{k^*}, b_k\ne b_{k^*}\},
\end{equation}
where clearly $\mathcal{E}^+ = \mathcal{E}_{aux}^+ \cup \mathcal{E}_{k^*}$. These constructed subgraphs provide the necessary paired data to train the feature alignment network in the subsequent adversarial module.

\subsubsection{Adversarial Distribution Alignment}
Based on the constructed view-specific subgraphs, we employ LightGCN~\cite{r22} as the backbone encoder to perform representation learning. This allows us to extract latent representations of users and items from distinct topological structures. Taking user embeddings in the positive context graph $\mathcal{G}^+$ as an example, the hierarchical propagation and aggregation mechanism is recursively calculated as follows:
\begin{equation}
\begin{aligned}
    \mathbf{e}_{u}^{+,(l+1)}&=\sum_{i\in\mathcal{N}_u^+}\frac{1}{\sqrt{|\mathcal{N}_u^+||\mathcal{N}_i^+|}}\mathbf{w}_{b}\circ\mathbf{e}_{i}^{+,(l)}, \\
    \mathbf{e}_u^+&=\frac{1}{L+1}\sum_{l=0}^{L}\mathbf{e}_u^{+,(l)},
\label{eq3}
\end{aligned}
\end{equation}
where $\mathbf{e}_{u}^+$ denotes the user embedding under subgraph $\mathcal{G}^+$, $L$ is the number of GCN layers, \(\mathcal{N}_{u}^+\) and \(\mathcal{N}_{i}^+\) represent the neighbor sets of user \(u\) and item \(i\), and $\circ$ denotes element-wise multiplication. To capture the distinct influence of different behavior types, we define the behavior-aware transformation vector $\mathbf{w}_{b}$ as:
\begin{equation}
\mathbf{w}_{b}=f\left(\mathop \| \limits_{k = 1}^K\mathbf{e}_{b_k}\right),
\end{equation}
where $\|$ is the concatenation operation, $\mathbf{e}_{b_k}$ is the learnable embedding of behavior $k$, and $f(\cdot)$ denotes the feature alignment network, which applies a nonlinear transformation over the concatenated behavior embeddings. After recursive aggregation, we obtain the user and item embeddings from the positive context subgraph $\mathcal{G}^{+}$, denoted as $\mathbf{e}_u^+, \mathbf{e}_i^+$, and from the auxiliary view subgraph $\mathcal{G}_{aux}^{+}$, denoted as $\mathbf{e}_{u,aux}^+$, $\mathbf{e}_{i,aux}^+$. 

Although $\mathcal{G}^+$ and $\mathcal{G}_{aux}^+$ share the same user and item nodes, they exhibit significant topological differences: the former encodes the ground truth of target service invocations, while the latter only captures exploratory auxiliary patterns. This structural discrepancy results in a distribution shift between the learned representations. To bridge this gap and enable the model to infer target intents from auxiliary signals, we introduce an adversarial domain adaptation mechanism. A discriminator $\mathcal{D}(\cdot)$ is trained to distinguish whether a given feature embedding originates from the target-rich domain ($\mathcal{G}^+$) or the auxiliary domain ($\mathcal{G}_{aux}^+$), while the feature alignment network $f(\cdot)$ is jointly optimized to fool the discriminator. This min-max game encourages the auxiliary-induced embeddings to align with the high-level semantics of the target behavior, effectively mitigating the domain discrepancy. The adversarial loss is defined as:
\begin{equation}
\mathcal{L}_{{adv}}=\mathbb{E}_{\mathcal{G}^+}[\log \mathcal{D}(\mathbf{e}^+)]+\mathbb{E}_{\mathcal{G}_{aux}^+}[\log(1-\mathcal{D}(\mathbf{e}_{aux}^+))],
\end{equation}
where $\mathbb{E}$ denotes the expectation operator, and $\mathbf{e}^+=\mathbf{e}^+_u\|\mathbf{e}^+_i$ and $\mathbf{e}_{aux}^+=\mathbf{e}_{u,aux}^+\|\mathbf{e}_{i,aux}^+$.

Upon convergence, the feature alignment network $f(\cdot)$ serves as a robust adapter that projects auxiliary behaviors into the target preference space. We then apply this learned alignment to the unlabeled interaction set (i.e., the auxiliary behavior subgraph without target invocations), denoted as $\mathcal{G}_{aux}^-=(\mathcal{V}_{aux}^-,\mathcal{E}_{aux}^-)$. This step is critical for false negative driscovery, as it allows us to evaluate the latent conversion potential of unobserved interactions. The edge set of $\mathcal{G}_{aux}^-$ is defined as:
\begin{equation}
    \mathcal{E}_{aux}^-=\{(u,i,b_k)\mid(u,i,b_{k^*})\not\in\mathcal{E}_{k^*},b_k\neq b_{k^*}\}.
\end{equation}
Similar to Eq.~(\ref{eq3}), we perform message passing over $\mathcal{G}_{aux}^-$ to obtain the enhanced user and item embeddings, denoted as $\mathbf{e}_{u,aux}^-$ and $\mathbf{e}_{i,aux}^-$, which are subsequently used to identify high-confidence false negatives. 

\subsubsection{Identifying High-Confidence False Negatives}
\label{sec1}

Based on the aligned representations obtained from the adversarial module, we aim to quantify the likelihood that an unobserved interaction in the auxiliary domain actually corresponds to a latent user need. We compute the predicted probability $y_{ui}^-$ for the target behavior $b_{k^*}$ via a multi-layer perceptron (MLP):
\begin{equation}
 y_{ui}^-=\mathrm{MLP}(\mathbf{e}_{aux}^-).
 \label{eq7}
\end{equation}
This score serves as a propensity estimate, indicating how structurally similar the auxiliary interaction is to a confirmed target invocation. Similarly, for the positive context subgraph $\mathcal{G}_{aux}^+$, the corresponding predicted scores $y_{ui}^{+}$ are computed according to Eq.~(\ref{eq7}). To guide the learning of this propensity estimator, we design a discriminative objective function. We treat the auxiliary views of confirmed target behaviors $\mathcal{V}_{aux}^+$ as positive anchors and the unobserved auxiliary interactions $\mathcal{V}_{aux}^-$ as tentative negatives:
\begin{equation}
\mathcal{L}_{score}=-\sum_{v\in\mathcal{V}_{aux}^+}\log y_{ui}^{+}-\sum_{v\in\mathcal{V}_{aux}^-}\log(1-y_{ui}^{-}).
\end{equation}
By jointly optimizing this supervised loss with the adversarial loss, the total alignment objective is defined as $\mathcal{L}_{align}=\mathcal{L}_{adv}+\mathcal{L}_{score}$. Crucially, the adversarial component $\mathcal{L}_{adv}$ prevents the model from overfitting to the tentative negative labels in $\mathcal{L}_{score}$ by forcing the feature distributions to align. Consequently, items in $\mathcal{G}_{aux}^-$ that share intrinsic semantics with $\mathcal{G}_{aux}^+$ will receive high predicted scores despite being labeled as zeros in $\mathcal{L}_{score}$. Finally, based on these refined propensity scores, we perform false negative discovery. We screen out auxiliary interactions that lack explicit target invocations but exhibit high conversion potential by retaining those with predicted scores greater than a threshold $\mu$. The items associated with these high-confidence interactions form the latent positive set, defined as:
\begin{equation}
\mathcal{E}_{lat} = \left\{ (u,i) \mid \exists (u,i,b_k)\in \mathcal{E}_{aux}^-,\ y_{ui}^{-} \geq \mu \right\}.
\end{equation}
These identified interactions are subsequently treated as recovered positive samples to augment the sparse target signal in the main recommendation task. The process of false negative discovery is detailed in Algorithm~\ref{alg1}. It is important to clarify that our False Negative Discovery mechanism does not rely solely on structural similarity within a single auxiliary view, which could indeed mistake ``comparison browsing'' for ``purchase intent''. Instead, NoVa mitigates this risk through cross-view distributional alignment. By integrating all types of auxiliary behaviors into the feature alignment network $f(\cdot)$, we project the comprehensive behavioral patterns into the target space. The adversarial loss forces the model to penalize auxiliary interactions that—despite being structurally connected—exhibit statistical distributions distinct from true target invocations. Therefore, the recovered positives in $\mathcal{E}_{lat}$ are not just high-degree nodes, but interactions that successfully fool the discriminator by statistically resembling the intrinsic preference distribution of the target behavior.

\begin{algorithm}[t]
\caption{False Negative Discovery}
\label{alg1}
\begin{algorithmic}[1]
\Require Multi-behavior graph $\mathcal{G}$, target behavior $b_{k^*}$, high-confidence false negatives threshold $\mu$.
\Ensure Latent positive set $\mathcal{E}_{lat}$.

\State  \textbf{Phase 1: View-Specific Subgraph Construction}
\State  Construct positive context subgraph $\mathcal{G}^+$ containing all interactions leading to target behavior.
\State  Construct auxiliary view subgraph $\mathcal{G}_{aux}^+$ from $\mathcal{G}^+$ by removing target edges to create a paired auxiliary view.
\State  Construct unlabeled auxiliary subgraph $\mathcal{G}_{aux}^-$ containing all auxiliary interactions without target behavior.

\State  \textbf{Phase 2: Adversarial Distribution Alignment}
\State  Initialize model parameters $\Theta$.
\While{not converged}
    \State \textbf{Forward Propagation:}
    \State Compute hierarchical embeddings $\mathbf{e}^+$, $\mathbf{e}_{aux}^+$, and $\mathbf{e}_{aux}^-$ via the LightGCN encoder.
    \State \textbf{Propensity Estimation:}
    \State Calculate propensity scores $y_{ui}^+$ and $y_{ui}^-$ using the feature alignment network.
    \State \textbf{Loss Computation:}
    \State Compute adversarial loss $\mathcal{L}_{adv}$ to align the distribution of $\mathcal{G}_{aux}^+$ with $\mathcal{G}^+$.
    \State Compute prediction loss $\mathcal{L}_{score}$ to evaluate the conversion potential of auxiliary behaviors.
    \State Calculate total alignment loss $\mathcal{L}_{align} = \mathcal{L}_{adv} + \mathcal{L}_{score}$.
    \State \textbf{Optimization:}
    \State Update parameters $\Theta$ by minimizing $\mathcal{L}_{align}$ via gradient descent.
\EndWhile

\State \textbf{Phase 3: Identifying High-Confidence False Negatives}
\State Initialize latent positive set $\mathcal{E}_{lat} = \emptyset$.
\For{each interaction $(u,i) \in \mathcal{E}_{aux}^-$}
    \State Compute prediction score $y_{ui}^- = \text{MLP}(\mathbf{e}_{u,aux}^- \| \mathbf{e}_{i,aux}^-)$.
    \If{$y_{ui}^- \geq \mu$}
        \State Add $(u,i)$ to $\mathcal{E}_{lat}$ as a recovered high-confidence false negative.
    \EndIf
\EndFor
\State \Return $\mathcal{E}_{lat}$
\end{algorithmic}
\end{algorithm}

\subsection{Label Noise Suppression}
\subsubsection{Behavior-Aware Representation Learning}
To comprehensively model user preference features under multi-behavior scenarios, based on LightGCN, we separately learn user and item embeddings $\mathbf{e}_u^k$ and $\mathbf{e}_i^k$ from each auxiliary behavior subgraph $\mathcal{G}_k$ ($k \neq k^*$) and the target subgraph $\mathcal{G}_{k^*}$. Recognizing that different auxiliary behaviors contribute unevenly to the target preference, we employ a multi-head Attention mechanism to capture the intrinsic cross-behavior dependencies. Specifically, for a behavior pair $(b_k, b_{k^*})$, we calculate the attention weights to quantify the contribution of the auxiliary behavior $b_k$. The attention score $\alpha(k,k^*)$ is computed as:
\begin{equation}
    \alpha(k,k^*)={\mathop \| \limits_{h = 1}^H }\beta_{k,k^*}^{h}\mathbf{W}_{V}^{h}\mathbf{e}_{i}^k,
\end{equation}
where \(H\) represents the number of heads in the multi-head attention mechanism indexed by \(h\), \({\bf{W}}_V^h\in\mathbb{R}^{d/H\times d}\) denotes the value embeddings under the head index \(h\), \(d\) is the embedding dimension, and \(\beta_{k,k^*}^h\) represents the \(h\)-th multi-head attention weight of the behavior pair, calculated as follows:
\begin{equation}
\begin{aligned}
\beta_{k,k^*}^h &= \frac{\exp(\bar{\beta}_{k,k^*}^h)}{\sum_{k\neq k^*} \exp(\bar{\beta}_{k,k^*}^h)}, \\
\bar{\beta}_{k,k^*}^h &= \frac{({\bf{W}}_Q^h \cdot {\bf{e}}_i^{k})^\top \cdot ({\bf{W}}_K^h \cdot {\bf{e}}_i^{k^*})}{\sqrt{d/H}},
\end{aligned}
\end{equation}
where \({\bf{W}}_Q^h\in\mathbb{R}^{d/H\times d}\) and \({\bf{W}}_K^h\in\mathbb{R}^{d/H\times d}\) represent the query and key embeddings. Based on these learned weights, we derive the initial transferred feature representation $m_{k\rightarrow k^*}$, which aggregates auxiliary knowledge into the target space:
\begin{equation}
m_{k\rightarrow k^*}=\alpha(k,k^*)\cdot\sigma(\mathbf{W}_{k\rightarrow k^*}({\mathbf{e}_i^k}^\top\cdot \mathbf{e}_{i}^{k^*})),
\end{equation}
where $\sigma(\cdot)$ is the sigmoid activation function and $\mathbf{W}_{k\rightarrow k^*}$ is a learnable transformation matrix. Crucially, while $m_{k\rightarrow k^*}$ captures the fused auxiliary information, it may still contain label noise inherited from ambiguous interactions. 

\subsubsection{Semantic-Guided Bias Correction}
While the attention mechanism effectively aggregates auxiliary features, simply fusing them implicitly assumes that all auxiliary interactions are beneficial. However, typical contrastive learning approaches, which enforce strict alignment between auxiliary and target views, are ill-suited for this scenario. The rationale is that auxiliary logs inherently contain label noise that does not exist in the target behavior distribution. Forcing the target representation to align with a noisy auxiliary view would inadvertently propagate this noise, degrading the purity of the target preference modeling.
 
Therefore, instead of blind alignment, we propose a subtraction-based semantic-guided bias correction strategy to explicitly filter out misleading signals. We first quantify the distributional bias introduced by non-converted interactions. We estimate the initial bias representation $\delta_{k\to k^*}^0$ as the centroid difference between active items (present in both behaviors) and ambiguous items (present only in auxiliary behavior):
\begin{equation}
\delta_{k\to k^*}^0=\mathbb{E}_{i\in\mathcal{I}_{k,k^{*}}^+}[\mathbf{e}_i^k]-\mathbb{E}_{i\in\mathcal{I}_{k,k^*}^-}[\mathbf{e}_i^k],
\end{equation}
where $\mathcal{I}_{k,k^{*}}^+$ and $\mathcal{I}_{k,k^{*}}^-$ denote the sets of overlapping and non-overlapping items, respectively. To prevent over-correction (i.e., filtering out valid exploratory interests), we refine this bias using service content similarity as a semantic gate. 
The intuition is straightforward: if an auxiliary service is semantically dissimilar to the services the user has definitively invoked, it is highly likely to be noise. We compute the semantic deviation $\delta_{k\to k^*}$ by scaling the initial bias with a similarity-based penalty:
\begin{equation}
\delta_{k\to k^*}=\left(1-\frac{1}{|\mathcal{I}_{k}|}\sum_{i\in\mathcal{I}_{k}}\max_{j\in\mathcal{I}_{k^*}}\mathrm{sim}(\mathbf{e}_i,\mathbf{e}_j)\right)\cdot\delta_{k\to k^*}^0,
\end{equation}
where $\mathrm{sim}(\cdot)$ denotes cosine similarity. Mathematically, the term in parentheses serves as a dynamic gating coefficient: low semantic similarity drives this value towards 1 to enforce bias subtraction, while high similarity minimizes it to prevent over-correction. Finally, we perform the feature integration with bias correction. We subtract the estimated semantic noise $\delta_{k\to k^*}$ from the aggregated auxiliary features $m_{k\to k^*}$ to obtain the purified transfer signal $\mathcal{T}_{ \to k^*}$. The final update rule for the target item representation $\mathbf{e}_i^{k^*,(l+1)}$ is formulated as:
\begin{equation}
\begin{aligned}
        \mathbf{e}_i^{k^*,(l+1)} &= \sum\limits_{u \in {{\mathcal N}_i}} \frac{1}{\sqrt{|{{\mathcal N}_u}||{{\mathcal N}_i}|}} \mathbf{e}_u^{k^*,(l)} + {\mathcal{T}_{ \to k^*}},\\
        \mathcal{T}_{ \to k^*}&=\sum_{k\neq k^*} (m_{k\to k^*} - \delta_{k\to k^*}).   
        \label{eq16}
\end{aligned}
\end{equation}
By explicitly subtracting $\delta_{k\to k^*}$, NoVa ensures that the model selectively absorbs valuable signals while suppressing irrelevant interference caused by distribution shifts. It is worth noting that this similarity-based filtering strategy aligns with the fundamental philosophy of CF. By penalizing interactions that deviate significantly from the user's established preference manifold, we effectively reinforce the preference consistency assumption inherent in CF, ensuring that the transferred auxiliary knowledge remains structurally compatible with the target service predictions.

\subsection{Joint Optimization}

After multiple layers of behavior-specific feature aggregation and the semantic-guided bias correction, we obtain the refined representations of user $u$ and item $i$ under the target behavior $b_{k^*}$. The final representation is derived by summing the embeddings across all $L$ layers:
\begin{equation}
\mathbf{e}_i^{k^*}=\sum_{l=0}^{L}\mathbf{e}_i^{k^*,(l)},\quad\mathbf{e}_u^{k^*}=\sum_{l=0}^{L}\mathbf{e}_u^{k^*,(l)},
\end{equation}
where $\mathbf{e}_u^{k^*, (l)}$ and $\mathbf{e}_i^{k^*, (l)}$ are obtained through the noise-resilient aggregation mechanism described in Eq.~(\ref{eq16}). 

To effectively incorporate the recovered high-confidence false negatives into the main service recommendation task, we adopt a weighted BPR-based loss~\cite{r23}. The objective is to optimize the model's ranking ability by maximizing the prediction score difference between observed (positive) and unobserved (negative) service pairs. Specifically, given a positive item $i$ and a sampled negative item $j$, the recommendation objective is defined as:
\begin{equation}
\mathcal{L}_{rec}=\sum_{(u,i,j)\in\mathcal{O}}\omega_{ui}\cdot(-\log\sigma(y_{u,i}^{k^*}-y_{u,j}^{k^*})),
\end{equation}
where \({y_{u,i}^{k^*}} = {{\mathbf{e}}_u^{k^*}}^\top{\mathbf{e}}_i^{k^*}\) denotes the predicted interaction score. The training set $\mathcal{O}$ is constructed by augmenting the original observed interactions with our discovered latent positives. We define the positive set as $\mathcal{O}^+ = \mathcal{O}_{true}^+ \cup \mathcal{O}_{lat}^+$. Here, $\mathcal{O}_{true}^+$ denotes the set of confirmed ground-truth interactions, and $\mathcal{O}_{lat}^+ = \mathcal{E}_{lat}$ represents the set of recovered latent positives identified by our adversarial feature alignment module.

From a positive-unlabeled learning perspective, recovered samples inherently carry higher uncertainty than ground-truth samples. 
To reflect this, we introduce a confidence coefficient $\omega_{ui}$ to differentiate their contributions during gradient updates:
\begin{equation}
\omega_{ui} = 
\begin{cases} 
1, & \text{if } (u,i) \in \mathcal{O}_{true}^+ \\
\beta, & \text{if } (u,i) \in \mathcal{O}_{lat}^+ 
\end{cases}
\end{equation}
where $0 < \beta < 1$ is a hyperparameter controlling the impact of the latent positive samples. 
This weighting strategy ensures that the model benefits from the augmented data while preventing overfitting to potentially noisy pseudo-labels.

Finally, we jointly optimize the recommendation task with the adversarial alignment task. 
The total objective function is formulated as:
\begin{equation}
\mathcal{L}_{total} = \mathcal{L}_{rec} + \lambda_1 \mathcal{L}_{align} + \lambda_2 \|\Theta\|_2^2,
\end{equation}
where $\Theta$ denotes the set of all model parameters. 
$\mathcal{L}_{align} = \mathcal{L}_{adv} + \mathcal{L}_{score}$ is the alignment loss defined in Section~\ref{sec1}, which guides the discovery of false negatives. 
$\lambda_1$ balances the trade-off between the main recommendation task and the auxiliary alignment task, while $\lambda_2$ serves as the coefficient for $L_2$ regularization to prevent overfitting.
\subsection{Further Analysis}
\subsubsection{Alignment Convergence}
To provide theoretical guarantees for our adversarial feature alignment module, we analyze the convergence properties of the min-max objective. Let $p_{target}(\mathbf{x})$ and $p_{aux}(\mathbf{x})$ denote the feature distributions of the positive context graph $\mathcal{G}^+$ and the auxiliary view $\mathcal{G}_{aux}^+$, respectively. The discriminator $D(\cdot)$ aims to distinguish between samples from these two distributions, while the alignment network $f(\cdot)$ aims to minimize the Jensen-Shannon divergence between them. The optimal discriminator $D^*(\cdot)$ for a fixed generator $G$ is given by:
\begin{equation}
D^*(\mathbf{x}) = \frac{p_{target}(\mathbf{x})}{p_{target}(\mathbf{x}) + p_{aux}(\mathbf{x})}.
\end{equation}
Substituting $D^*$ into the adversarial loss $\mathcal{L}_{adv}$, the objective function for the generator becomes minimizing:
\begin{equation}
\mathcal{L}_{G} = 2 \cdot D_{JS}(p_{target} \| p_{aux}) - 2 \log 2,
\end{equation}
where $D_{JS}$ represents the Jensen-Shannon divergence. Since $D_{JS} \geq 0$ and equals zero if and only if $p_{target} = p_{aux}$, the global minimum of the training criterion is achieved when the auxiliary feature distribution perfectly matches the target distribution. At this equilibrium, the alignment network $f(\cdot)$ has successfully learned to project auxiliary interactions into the target semantic space, ensuring that the subsequently identified high-confidence false negatives in $\mathcal{G}_{aux}^-$ are statistically indistinguishable from true positive service invocations. The optimal discriminator $D^*(x)$ estimates the density ratio between the target distribution $p_{target}$ and the auxiliary distribution $p_{aux}$.Consequently, the propensity score $y_{ui}^-$ produced by the alignment network serves as a proxy for the conditional probability that an unlabeled sample originates from the latent target distribution. Setting a threshold $\mu$ is effectively equivalent to performing a likelihood ratio test: we only accept auxiliary samples where the probability of being a ``latent target'' outweighs the probability of being ``noise'' by a margin defined by $\mu$. Although a closed-form bound relative to the true positive prior is intractable due to the unobserved nature of PU learning, our sensitivity analysis in Section~\ref{sec5d} confirms that the model's performance follows a predictable pattern relative to data density, validating $\mu$ as a robust control for the false positive rate.

\subsubsection{Gradient Perspective on Noise Suppression}
We further analyze how the proposed semantic-aware filtering mitigates negative transfer from a gradient optimization perspective. Consider the parameter update for the target item embedding $\mathbf{e}_i^{k^*}$. Without bias correction, the gradient descent update rule would be driven by a mixture of valid signals and noise:
\begin{equation}
\nabla \mathcal{L} \propto \frac{\partial \mathcal{L}_{rec}}{\partial \mathbf{e}_i^{k^*}} + \sum_{k \neq k^*} \gamma_k \frac{\partial \mathcal{L}_{aux}}{\partial \mathbf{e}_i^k},
\end{equation}
where the second term represents the influence of auxiliary behaviors. In the presence of label noise, the auxiliary gradient $\nabla_{noise}$ may point in a direction orthogonal or even opposite to the true target gradient $\nabla_{target}$ (i.e., $\langle \nabla_{noise}, \nabla_{target} \rangle < 0$), leading to oscillation or sub-optimal convergence.

By explicitly subtracting the semantic bias term $\delta_{k \to k^*}$ in Eq.~\ref{eq16}, our method effectively acts as a gradient rectifier. The bias term $\delta_{k \to k^*}$, which captures the centroid difference between confirmed and ambiguous items, estimates the expected direction of the noise gradient. Subtracting this component ensures that the propagated auxiliary gradient is projected onto the subspace consistent with the target semantics, thereby guaranteeing that $\langle \nabla_{aux}^{rectified}, \nabla_{target} \rangle \geq 0$. This theoretical property ensures robust optimization even when the auxiliary data is heavily corrupted by distribution shifts.

\section{Experiment}

In this section, we conduct extensive experiments on three real-world datasets to answer the following research questions: 
\begin{itemize}
    \item \textbf{RQ1:} How does the performance of NoVa compare with various state-of-the-art recommendation models? 
    \item \textbf{RQ2:} What are the specific contributions of the key components in NoVa to its overall performance? 
    \item \textbf{RQ3:} How does NoVa perform under different hyperparameter settings?
    \item \textbf{RQ4:} Can NoVa truly detect false positive samples?
\end{itemize}
\subsection{Experimental Settings}
\subsubsection{Datasets}

\begin{table}[t]
\caption{Datasets Statistics}
\label{tab1}
\centering
\footnotesize
\renewcommand{\arraystretch}{1.3}
\begin{tabular}{
    >{\centering\arraybackslash}m{0.6cm} |
    >{\centering\arraybackslash}m{0.7cm}
    >{\centering\arraybackslash}m{0.7cm}
    >{\centering\arraybackslash}m{0.8cm}
    >{\centering\arraybackslash}m{3.4cm}
}
    \toprule
    Stats. & \#Users & \#Items & \#Interactions & \#Behavior Type\\ \hline
    Tmall & 31,882 & 31,232 & 1,451,219 & \{View, Fav., Cart, Purchase\}\\ 
    IJCAI & 17,435 & 35,920 & 799,368 & \{View, Fav., Cart, Purchase\}\\
    Retail & 2,174 & 30,113 & 97,381 & \{View, Cart, Purchase\}\\ 
    \hline
\end{tabular}
\end{table}

\begin{table}[t]
    \centering
    \small
    \caption{Performance Comparison of Various Methods}
    \renewcommand{\arraystretch}{1.2}
    \setlength{\tabcolsep}{1mm}
    \begin{tabular}{c|cc|cc|cc}
    \hline
    \multirow{2}*{Model} & \multicolumn{2}{c|}{Tmall} & \multicolumn{2}{c|}{IJCAI-Contest} & \multicolumn{2}{c}{Retail Rocket} \\ \cline{2-7}
     & HR & NDCG & HR & NDCG & HR & NDCG \\ \hline
    BPR & 0.3258 & 0.1560 & 0.2716 & 0.1115 & 0.2247 & 0.1341 \\
    LightGCN & 0.3723 & 0.2079 & 0.3442 & 0.1512 & 0.2514 & 0.1522 \\ \hline
    MBGCN & 0.4602 & 0.2628 & 0.3919 & 0.1868 & 0.3036 & 0.1837 \\
    CML & 0.5074 & 0.3087 & 0.4308 & 0.2371 & 0.3141 & 0.1985 \\
    KHGT & 0.5226 & 0.3284 & 0.4533 & 0.2637 & 0.3209 & 0.2043 \\
    KMCLR & 0.5573 & 0.3532 & 0.4746 & 0.2801 & 0.3325 & 0.2118 \\ \hline
    NMTR & 0.4619 & 0.2653 & 0.3917 & 0.1852 & 0.3092 & 0.1876 \\
    COPF & 0.5486 & 0.3396 & 0.4758 & 0.2744 & 0.3511 & 0.2264 \\
    MPC & 0.5601 & 0.3519 & 0.4820 & 0.2816 & 0.3480 & 0.2439 \\ \hline
    MixGCF & 0.5539 & 0.3484 & 0.4897 & 0.2811 & 0.3475 & 0.2498 \\ 
    DeMBR & \underline{0.5926} & \underline{0.3712} & \underline{0.5132} & \underline{0.3189} & \underline{0.3721} & \underline{0.2549} \\ \hline
    \textbf{NoVa} & \textbf{0.6205} & \textbf{0.3911} & \textbf{0.5547} & \textbf{0.3396} & \textbf{0.3927} & \textbf{0.2726} \\
    \textbf{\%Improv.} & \textbf{4.71} & \textbf{5.36} & \textbf{8.09} & \textbf{6.49} & \textbf{5.54} & \textbf{6.94} \\ 
    \textbf{\textit{p}-value} & $4e^{-3}$ & $1e^{-2}$ & $9e^{-4}$ & $2e^{-3}$ & $3e^{-2}$ & $1e^{-3}$ \\ \hline
    \end{tabular}    
\label{tab2}
\end{table}

We conduct a systematic evaluation of the proposed NoVa model on three real-world datasets. To ensure fair comparisons, all experiments are based on publicly available datasets preprocessed following the protocol of~\cite{r17}. The statistical information of the datasets is summarized in Table~\ref{tab1}: 
\begin{itemize}
    \item \textbf{Tmall\footnote{\url{https://tianchi.aliyun.com/dataset/140281}.}:} This dataset is collected from a major Chinese e-commerce platform and records four common types of user behaviors during the shopping process: \textit{view}, \textit{favorite}, \textit{add-to-cart}, and \textit{purchase}. Among them, \textit{purchase} is regarded as the target behavior, while the other three are treated as auxiliary behaviors.
    \item \textbf{IJCAI-Contest\footnote{\url{https://tianchi.aliyun.com/dataset/42}.}:} This dataset originates from the IJCAI 2015 Challenge and captures retail behaviors of companies toward customers. It shares the same behavior types as the \textbf{Tmall} dataset.
    \item \textbf{Retail Rocket\footnote{\url{https://www.kaggle.com/datasets/retailrocket/ecommerce-dataset}.}:} This dataset is collected from a real-world e-commerce platform, covering a 4.5-month span of user behavior logs. It includes three major behaviors: \textit{view}, \textit{add-to-cart}, and \textit{purchase}, which can effectively reflect the evolution of user preferences throughout the shopping journey and \textit{purchase} is regarded as the target behavior.
\end{itemize}

\subsubsection{Baselines}

We evaluate the performance of the NoVa model by comparing it with the following methods. To ensure fairness, we adopt the same hyperparameter settings as specified in their original publications and official implementations:
\begin{itemize}
    \item \textbf{BPR}~\cite{r23} introduces a principled criterion for learning from implicit feedback by directly optimizing for personalized ranking.
    \item \textbf{LightGCN}~\cite{r22} streamlines graph convolutional networks by retaining only the neighbor aggregation step.
    \item \textbf{MBGCN}~\cite{r6} constructs a unified multi-behavior graph and designs a two-layer propagation mechanism.
    \item \textbf{CML}~\cite{r17} employs a contrastive learning mechanism to distill transferable knowledge across different behavior.
    \item \textbf{KHGT}~\cite{r8} integrates a hierarchical graph structure with temporal encoding and graph attention mechanisms.
    \item \textbf{KMCLR}~\cite{r26} incorporates a multi-behavior learning module to capture users’ personalized preferences and a knowledge enhancement module to improve item representations.
    \item \textbf{NMTR}~\cite{r7} jointly models user behaviors through multi-task learning and explicitly captures the cascade relationship between behaviors.
    \item \textbf{COPF}~\cite{r21} formulates behavior fusion as a combinatorial optimization problem, applying constraint-based modeling at each behavior stage to improve behavior fusion. 
    \item \textbf{MPC}~\cite{r13} constructs multiple purchase chains to better model behavior patterns and builds a negative graph to filter out misleading interactions.
    \item \textbf{MixGCF}~\cite{d14} proposes a hop-mixing technique to synthesize hard negative samples by aggregating embeddings from different graph layers.
    \item \textbf{DeMBR}~\cite{r11} utilizes a memory pruning mechanism combined with semantic guidance to explicitly filter out noise from user interactions.
\end{itemize}

\subsubsection{Evaluation}
\begin{figure}[t]
    \centering
    \includegraphics[width=0.95\linewidth]{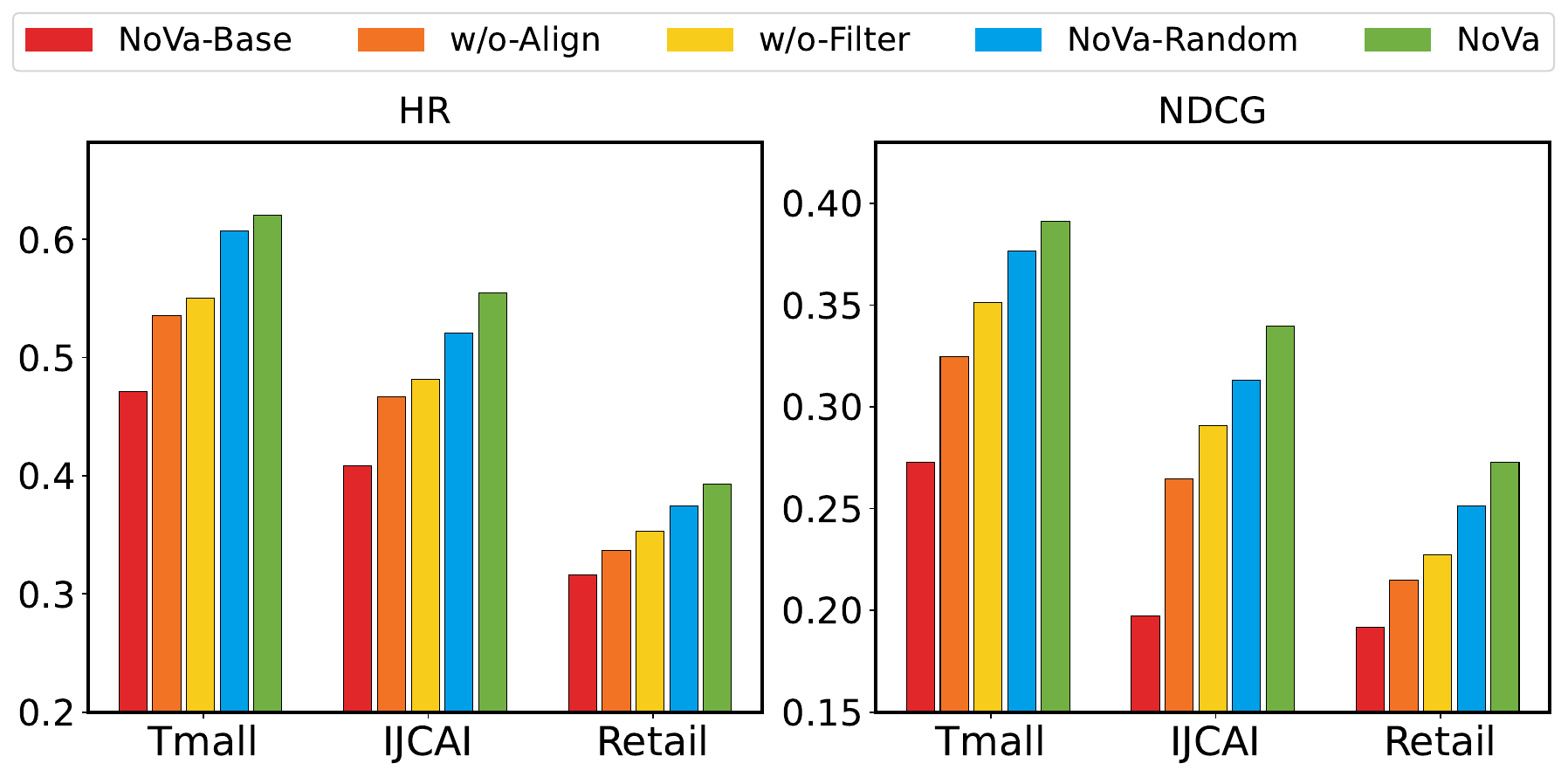}
    \caption{Ablation studies of NoVa, where IJCAI stands for IJCAI-Contest, and Retail stands for Retail Rocket.}
    \label{fig3}
\end{figure}
We adopt the widely used leave-one-out evaluation strategy~\cite{a5} to simulate the real-world recommendation scenario. For each user, the most recent interaction under the target behavior is held out as the test instance, while the remaining interactions are used for training. To rigorously assess the recommendation performance, we employ the all-ranking strategy, which ranks the ground-truth item against all unobserved items rather than a sampled subset. Performance is measured using two representative metrics: HR@K, which measures whether the target item is present in the top-$K$ list, and NDCG@K, which accounts for the position of the hit by assigning higher scores to items ranked higher. We report results with $K=10$. To ensure experimental robustness, we repeat the training and evaluation process 10 times for both NoVa and the best-performing baselines. We report the average results and perform a paired $t$-test to verify statistical significance, with the confidence level set at 95\% (i.e., $p < 0.05$).

\subsubsection{Parameters Settings}

We implement the proposed NoVa model using the PyTorch framework. The embedding dimension is fixed at $d=64$, and all parameters are initialized using the Xavier method~\cite{r24} to ensure a uniform distribution. We optimize the model using the Adam~\cite{r25} optimizer with a batch size of 2048 and a learning rate of $1e^{-3}$. The number of GCN propagation layers $L$ is tuned in the range of $\{1, 2, 3, 4\}$. Regarding the hyperparameters specific to NoVa, we explicitly set the alignment weight $\lambda_1$ to $0.1$ to balance the adversarial learning. The false negative discovery threshold $\mu$, which determines the selection of high-confidence latent positives, is tuned in the range of $\{0.2, 0.4, 0.6, 0.8, 1.0\}$. To regulate the contribution of these recovered latent samples, the confidence coefficient $\beta$ is searched within $\{0.1, 0.3, 0.5, 0.7, 0.9\}$. Furthermore, the $L_2$ regularization coefficient $\lambda_2$ is fixed at $1e^{-4}$ to prevent overfitting. 

\subsection{Performance Comparison (RQ1)}

Table~\ref{tab2} presents the performance comparison between NoVa and all baseline methods across the three datasets. The best results are highlighted in bold and the second-best results are underlined. \%Improv. represents the improvement of NoVa compared to the second-best model. Based on the results, we make the following observations: 

\begin{itemize}
    \item NoVa consistently achieves the best performance across all datasets. Specifically, NoVa yields performance improvements of 5.04\%, 7.29\%, and 6.24\% on Tmall, IJCAI-Contest, and Retail Rocket, respectively, compared to the best baseline. Rigorous paired $t$-tests confirm that these improvements are statistically significant. Among the compared models, single-behavior methods (BPR, LightGCN) perform the worst. This indicates that in sparse service ecosystems, relying solely on target service invocations is insufficient. Incorporating auxiliary behavioral signals is essential for building a comprehensive understanding of user service preferences.

    \item Compared to robust baselines, traditional multi-behavior models (e.g., MBGCN, CML, NMTR) exhibit relatively limited performance. A critical limitation is that these models typically adopt a closed-world assumption, indiscriminately treating all non-converted interactions as negative samples. They overlook the significant distribution shift between exploratory auxiliary behaviors and decisive target behaviors. Such coarse-grained modeling fails to align the feature distributions, leading to negative transfer effects where the distinct patterns of auxiliary noise misguide the target optimization process.

    \item Advanced robust methods, such as MixGCF and DeMBR, show competitive performance by incorporating negative sampling optimization or noise pruning mechanisms. DeMBR, in particular, effectively suppresses label noise via memory pruning. However, NoVa still outperforms DeMBR significantly. We attribute this to a fundamental difference in philosophy: while DeMBR primarily focuses on filtering out noise, it risks discarding ambiguous interactions that are actually high-confidence false negatives. In contrast, NoVa adopts a positive-unlabeled learning perspective. By employing adversarial feature alignment, it not only filters noise but actively recovers latent positive samples from the unlabeled data. This result confirms that uncovering value from weak signals is as critical as simple denoising in sparse recommendation.
\end{itemize}

\subsection{Ablation Study (RQ2)}

To systematically evaluate the contribution of each component in NoVa, we compare the full model against four variants:
\begin{itemize}
    \item \textbf{NoVa-Base:} It removes both the adversarial feature alignment and semantic-aware filtering modules. It retains only the multi-head attention mechanism to fuse explicit auxiliary features.
    \item \textbf{w/o-Align:} In this variant, the false negative discovery mechanism is removed. The model trains only on observed target interactions without recovering any latent positive samples from the auxiliary data.
    \item \textbf{w/o-Filter:} This variant removes the semantic-aware filtering mechanism. It directly fuses aligned auxiliary features without applying the semantic consistency constraint to suppress label noise.
    \item \textbf{NoVa-Random:} This variant replaces the adversarial scoring mechanism with a random strategy. Instead of selecting high-confidence samples, it randomly selects the same number of auxiliary interactions as latent positives for training.
\end{itemize}
\begin{figure}[t]
    \centering
    \includegraphics[width=0.95\linewidth]{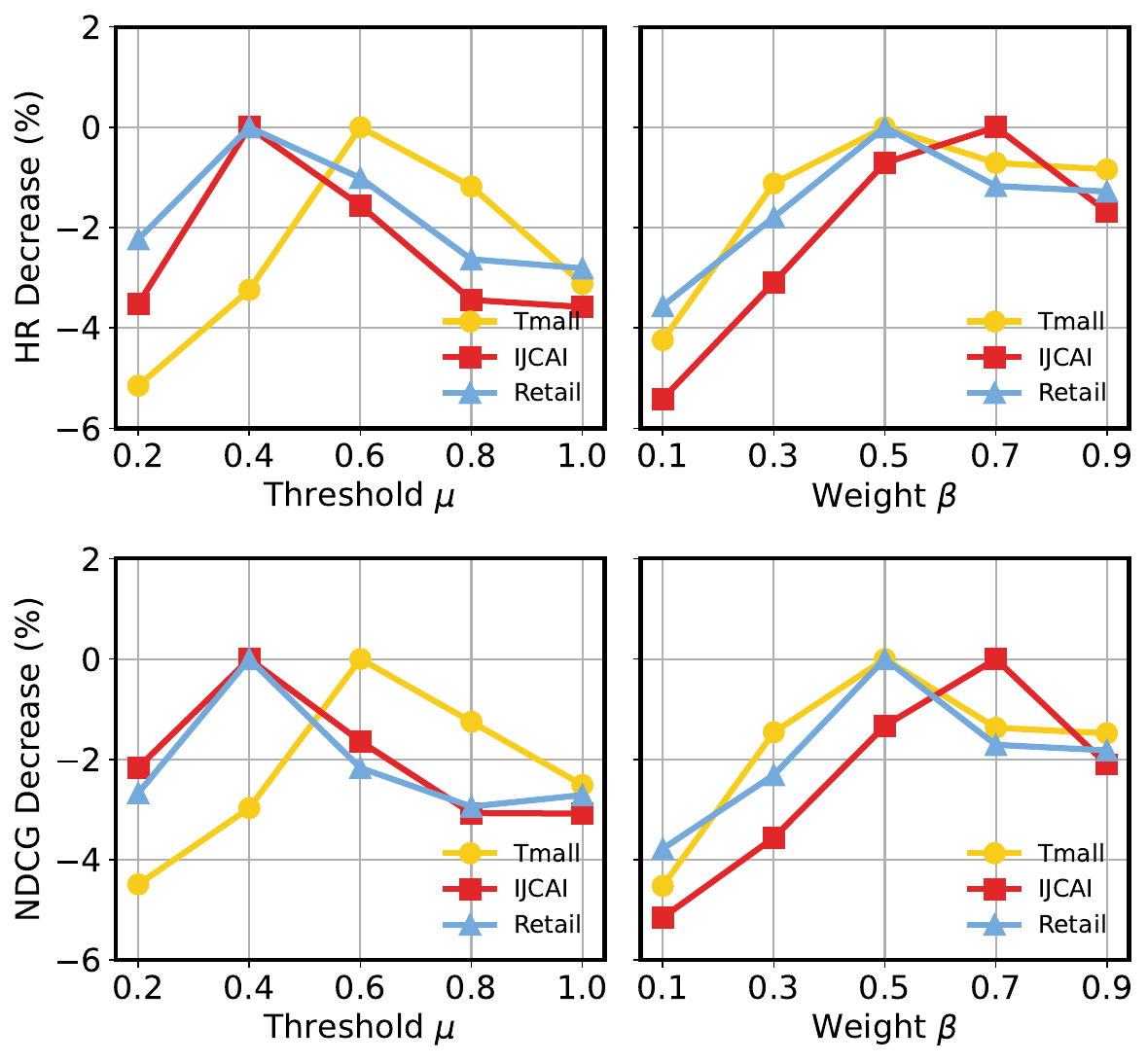}
    \caption{Hyperparameter $\mu$ and $\beta$ analysis of NoVa.}
    \label{fig4}
\end{figure}

Fig.~\ref{fig3} presents the performance comparison of all ablation variants across three datasets. We summarize the key observations as follows: (1) Removing the adversarial feature alignment module leads to a substantial performance drop. This result validates our core premise: the sparse target data alone is insufficient for training robust recommenders. By treating all non-converted interactions as negatives, w/o-Align discards valuable latent user intents. The superiority of the full NoVa model confirms that actively recovering high-confidence false negatives via distribution alignment effectively alleviates data sparsity and corrects sample selection bias. (2) The performance degradation of w/o-Filter highlights the risk of blind feature fusion. Without the semantic-aware filtering, the model becomes vulnerable to label noise inherent in web service logs. This noise introduces negative transfer, confusing the model's decision boundary. The full model's gain demonstrates that the semantic consistency constraint acts as an effective gatekeeper, ensuring that only high-quality, relevant auxiliary signals contribute to the target prediction. (3) Crucially, NoVa-Random performs worse than the full model. This provides compelling empirical evidence that the performance gains of NoVa are not merely due to data augmentation. Instead, it proves that our adversarial feature alignment module successfully identifies structurally valid latent needs. Randomly treating auxiliary behaviors as positives introduces massive false positives, which poisons the training process, whereas NoVa's selective strategy successfully turns noise into value. (4) Finally, NoVa-Base exhibits the worst overall performance. This indicates that the two proposed modules are complementary: ``alignment'' increases the recall of potential needs, while ``filtering'' ensures the precision of the utilized signals. Only by jointly deploying both can the model achieve robust service recommendation in sparse and noisy environments.

\subsection{Hyperparameter Analysis (RQ3)}
\label{sec5d}
\begin{table}[t]
\centering
\small
\caption{The Impact of The Number of GNN Layers $L$.}
\renewcommand{\arraystretch}{1.2}
\setlength{\tabcolsep}{1mm}
\begin{tabular}{c|cc|cc|cc}
\hline
    \multirow{2}*{Datasets} & \multicolumn{2}{c|}{Tmall} & \multicolumn{2}{c|}{IJCAI-Contest} & \multicolumn{2}{c}{Retail Rocket} \\ \cline{2-7}
     & HR & NDCG & HR & NDCG & HR & NDCG \\ \hline
    $L=1$ & 0.5724 & 0.3531 & 0.5166 & 0.2825 & 0.3369 & 0.2316 \\
    $L=2$ & 0.6132 & 0.3819 & 0.5349 & 0.3240 & 0.3691 & 0.2533 \\ 
    $L=3$ & \textbf{0.6205} & \textbf{0.3911} & \textbf{0.5547} & \textbf{0.3396} & \textbf{0.3927} & \textbf{0.2726} \\ 
    $L=4$ & 0.6017 & 0.3727 & 0.5331 & 0.3248 & 0.3752 & 0.2524 \\ 
    \hline
\end{tabular}
\label{tab3}
\end{table}
To evaluate the robustness of NoVa and the impact of key hyperparameters, we conduct a sensitivity analysis on two core parameters: the false negative discovery threshold $\mu$ and the confidence coefficient $\beta$. The results are illustrated in Fig.~\ref{fig4}.

The threshold $\mu$ determines the strictness of identifying high-confidence false negatives. A higher $\mu$ implies a more conservative selection strategy, while a lower $\mu$ allows for broader inclusion of latent positives. Experimental results indicate that the optimal $\mu$ is correlating with the interaction density of the dataset. For instance, on the dense Tmall dataset, a moderate threshold is preferred to sufficiently capture the abundant latent preferences without introducing excessive noise. Conversely, for datasets with different noise profiles, the threshold must be adjusted to balance the trade-off between false negative recall and false positive suppression. If $\mu$ is too small, the model may aggressively include noisy interactions as positive signals. If $\mu$ is too large, the model fails to recover enough latent signals, degenerating into a standard supervised model limited by data sparsity.

The coefficient $\beta$ controls the contribution weight of the recovered latent positives in the recommendation loss. For high sparsity scenario (IJCAI-Contest), the best performance is achieved at a relatively high weight ($\beta=0.7$). This suggests that in highly sparse settings where ground-truth signals are scarce, the model must rely heavily on the recovered latent positives to learn meaningful representations. The benefit of alleviating sparsity outweighs the risk of potential label noise, necessitating a more aggressive exploitation strategy. However, for data-rich scenario (Tmall/Retail Rocket), performance stabilizes around a moderate weight ($\beta=0.5$). In these scenarios, ground-truth labels are sufficient to guide the main optimization direction. Therefore, a balanced $\beta$ is optimal to treat the recovered samples as supplementary regularization rather than primary supervision, preventing the model from overfitting to pseudo-labels.

We further investigate the impact of the number of graph propagation layers $L$ in the LightGCN encoder, which controls the receptive field of the user and item embeddings. Table~\ref{tab3} summarizes the experimental results across all datasets. As observed, the performance improves significantly as $L$ increases from 1 to 3, achieving the optimal performance at $L=3$. This trend underscores the importance of modeling high-order connectivity in service recommendation. In sparse service ecosystems, first-order neighbors (direct interactions) are often insufficient to fully characterize user preferences. By stacking more propagation layers, NoVa effectively aggregates collaborative signals from multi-hop neighbors, enriching the semantic representation of users and services. However, further increasing $L$ to 4 leads to a performance degradation across all datasets. We attribute this drop to two primary factors: (1) Excessive propagation causes the embeddings of different nodes to become indistinguishable, reducing the model's ability to discriminate between specific user preferences. (2) Although NoVa incorporates a semantic-aware filtering mechanism, an overly deep architecture may aggressively aggregate irrelevant information from distant nodes. This introduces remote noise that outweighs the information gain, thereby confusing the decision boundary for false negative discovery. Consequently, we set $L=3$ as the default setting to balance efficient signal aggregation and noise robustness.

\subsection{Visualization of Recovered Latent Signals (RQ4)}
\begin{figure}[t]
    \centering
    \includegraphics[width=0.95\linewidth]{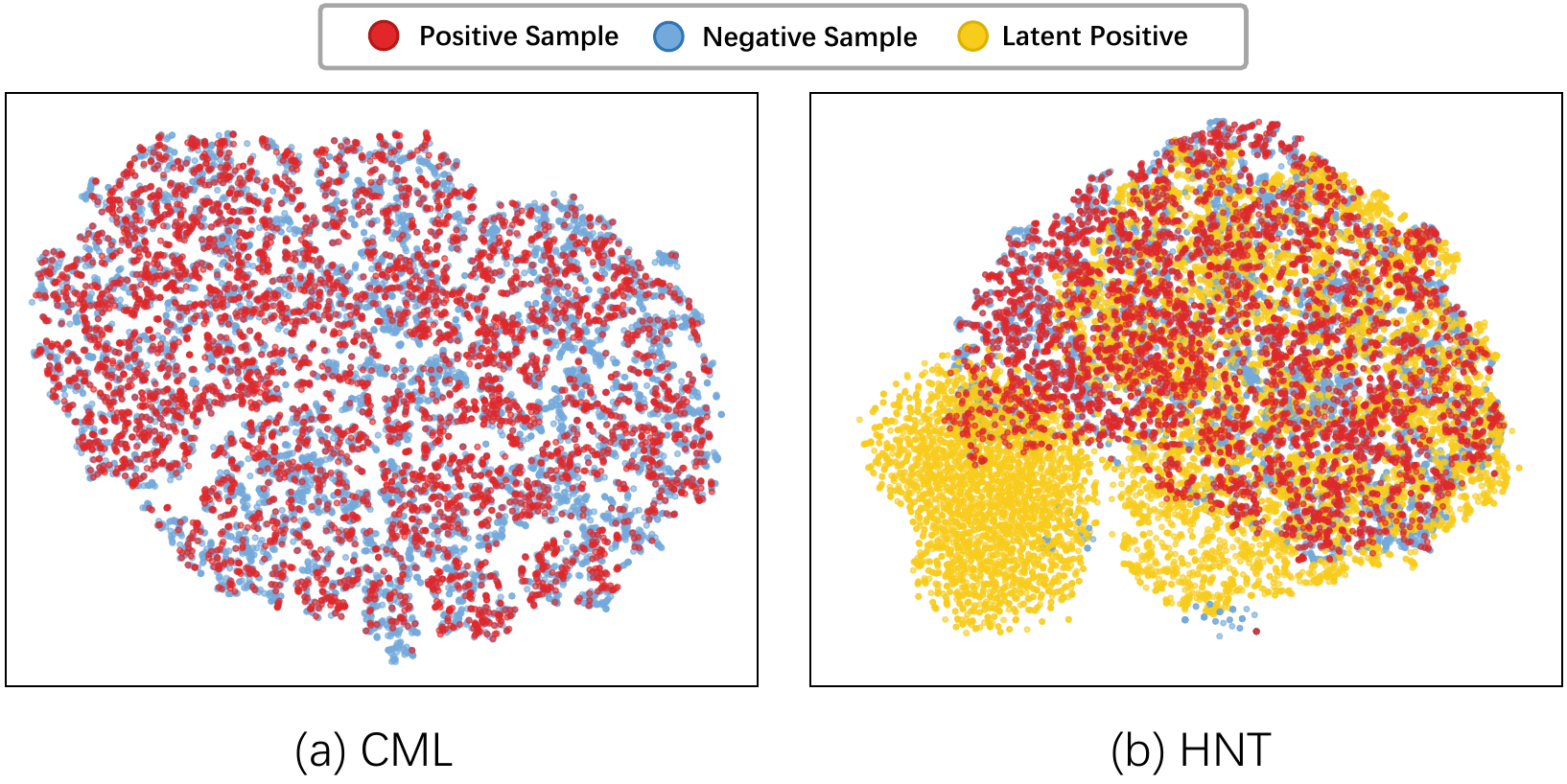}
    \caption{Visualization of sample distributions in the embedding space. The yellow points represent the latent positive set recovered by NoVa, illustrating how they bridge the distribution gap between observed positives and negatives.}
    \label{fig7}
\end{figure}
To intuitively validate whether NoVa successfully bridges the distribution gap and recovers meaningful signals, we visualize the learned embedding space. We employ t-SNE to project the user-service interaction embeddings into a 2-dimensional space. We select IJCAI because it exhibits the highest sparsity among all datasets (as detailed in Table 1). In such a data-scarce scenario, the distribution gap between observed positives and negatives is most pronounced, making the bridging effect of recovered latent signals most intuitively observable. Furthermore, we choose CML as the comparative baseline. Since CML operates under the standard closed-world assumption, it serves as an ideal reference to highlight the limitations of traditional alignment approaches. Fig.~\ref{fig7} compares the sample distributions of CML and our NoVa model, distinguishing three types of samples: observed positives (red), sampled negatives (blue), and the latent positives (yellow) recovered by NoVa.

As shown in Fig.~\ref{fig7}(a), the baseline CML exhibits a rigid separation structure. Although it distinguishes positives from negatives, there is a noticeable semantic vacuum (blank area) between the red and blue clusters. CML treats all unobserved interactions as hard negatives, forcing the model to push potentially ambiguous signals far away from the positive cluster. This results in a sharp, discontinuous decision boundary that lacks the capacity to capture latent or evolving user needs. In stark contrast, NoVa effectively populates this gap with the recovered latent prositives (yellow points). Crucially, these yellow points naturally form a bridge connecting the high-confidence positive region and the negative region. This distribution pattern provides strong empirical evidence that the adversarial module has successfully aligned the auxiliary features with the target space, effectively pulling the previously isolated ambiguous auxiliary interactions closer to the positive manifold. Consequently, these recovered false negatives smooth the transition between dislike and like, creating a continuous preference manifold rather than a disjointed decision boundary. Furthermore, the distinct cluster of yellow points extending outward likely represents exploratory interests. This confirms that NoVa possesses the generalization ability to discover novel service demands beyond mere memorization.

\section{Conclusion}

In this paper, we propose NoVa, a robust service recommendation framework that re-examines multi-behavior learning from a positive-unlabeled learning perspective. Unlike traditional approaches that rely on the closed-world assumption, we argue that auxiliary service logs are a mixture of high-confidence false negatives and label noise. Existing methods often overlook this distinction, leading to sample selection bias and negative transfer. To bridge this gap, NoVa employs a dual-mechanism strategy: (1) an adversarial feature alignment module that bridges the distribution difference between auxiliary and target behaviors to actively recover latent positive samples; and (2) a semantic-aware filtering module that suppresses label noise via service content consistency constraints. Extensive experiments on three real-world e-commerce datasets demonstrate that NoVa consistently outperforms state-of-the-art baselines, verifying its effectiveness in turning noise into value for sparse service recommendation. In future work, we plan to extend NoVa to support streaming service logs by incorporating temporal dynamics into the alignment process, further enhancing the real-time adaptability and interpretability of web service recommender systems.
\bibliographystyle{IEEEtran}
\bibliography{ref}

\vfill

\end{document}